# Matrix Mechanics Mis-Prized: Max Born's Belated Nobelization

John L. Heilbron[1+] and Carlo Rovelli[2,3,4*]

[1]Department of History, UC Berkeley, 3229 Dwinelle Hall, Berkeley, 94720-2550, California, USA.
[2]CPT, Aix-Marseille University, Université de Toulon, CNRS, Street, City, 13288, France.
[3]Department of Philosophy and the Rotman Institute of Philosophy, Western Ontario University, 1151 Richmond St, London, N6A5B7, Ontario, Canada.
[4] Perimeter Institute, 31 Caroline Street N, Waterloo, N2L2Y5, Ontario, Canada.

* E-mail: crovelli@uwo.ca.
+E-mail: johnheilbron@berkeley. Mailing address, April House, Shilton, OX18 4AB, UK.

## Abstract

We examine evaluations of the contributions of Matrix Mechanics and Max Born to the formulation of quantum mechanics from Heisenberg's Helgoland paper of 1925 to Born's Nobel Prize of 1954. We point out that the process of evaluation is continuing in the light of recent interpretations of the theory that deemphasize the importance of the wave function.

Keywords: Quantum mechanics, Matrix Mechanics, Probability, Nobel Prize.

## 1. Introduction

In 1954 Max Born received the Nobel prize for the probabilistic interpretation of the wave function he put forward in 1926[1]. The motivation for the belated prize should be as surprising to a modern physicist as the award to Einstein for the formula for the photo effect. For just as the Nobel citation omitted Einstein's theories of relativity, so it omitted Born's most significant contributions: recognition of matrix algebra in Heisenberg's cloudy breakthrough "Helgoland " paper of July 1925[2]; divination of the oracular principle $\mathbf{pq} -$ $\mathbf{qp} = \frac{h}{2\pi i}\mathbf{1}$ in the fundamental 1925 paper with Jordan[3]; and quick development of a full quantum mechanics [QM] based on it in the subsequent "three men" paper[4] . The "Matrix mechanics [MM]" of Born, Jordan, and Heisenberg and its application to the hydrogen atom by Pauli[5] were in hand before Schrödinger began to publish his widely influential papers on "Wave Mechanics [WM]" in January 1926.[6] Why did the Nobel establishment not recognize Born's contributions more appropriately? And why did it take so long to arrive at its inadequate characterization of them?

---

[1] Born, "Zur Quantenmechanik der Stoßvorgänge" (1926).
[2] Heisenberg, "Über quantentheoretische Umdeutung kinematischer und mechanischer Beziehungen" (1925).

This paper originates from these questions, which one of the authors (CR), a theoretical physicist engaged in current discussions of the interpretation of QM, posed to the other (JH), a historian interested in turning points in the description and understanding of the physical word.

There are good reasons why the judges in Stockholm rewarded Born how and when they did. But these good reasons contain a considerable burden of contingency in personality, timing, and institutions and may fail to bring out, and perhaps even disguise, the fundamental ideas of QM and Born's contributions to them. By teasing out Born's ideas, we can point up the contingent elements in the mixture of MM and WM that Born put together in his probabilistic interpretation of 1926 and that Bohr worked into his complementarity and his disciples developed into the "Copenhagen interpretation." Most of the scientific community came gradually to accept some version of the Copenhagen view about quantum physics, which the Nobel committee's final, favorable evaluation of Born's contributions confirmed.

The various Copenhagen interpretations including Bohr's complementarity have not closed discussions aimed at reaching a more satisfying understanding of the quantum revolution. The success of QM in domains like cosmology and astrophysics, remote from the laboratory context where Bohr's division between microsystem and measurement apparatus applies naturally; the discovery of the Bell inequalities and their experimental violations, recently recognized by a Nobel Prize; and a renewed interest in the conceptual questions left open by the Copenhagen synthesis have led to a renaissance of debates about the meaning of quantum mechanics. As opinions evolve, so does, or should, evaluation of the work of its founders, now over a century old. The process of assessment that reevaluated Born's contributions is still under way.

---

[3] Born and Jordan, "Zur Quantenmechanik" (1925).
[4] Born, Jordan, Heisenberg, "Zur Quantenmechanik II" (1926).
[5] Pauli, "Über das Wasserstoffspektrum vom Standpunkt der neuen Quantenmechanik" (1926).
[6] Schrödinger, "Quantisierung als Eigenwertproblem" (1926).

The key to the present discussions is recognition of the contingent and perturbative role of Schrödinger's wave mechanics. While interpretations that assign a heavy ontological status to the quantum state, such as Many Worlds[7] or Bohmian mechanics[8], see WM as the key conceptual novelty introduced in 1925/6, interpretations that devalue the ontological status of the quantum state, such as QBism[9], Healey's pragmatism[10], perspectival[11] or relational interpretations[12], make MM the centerpiece in the discovery of quantum mechanics. From this point of view, Born's claim in his Nobel acceptance speech to have "discover[ed] the foundations of a new way of thinking about the phenomena of nature" is not hyperbole and his probabilistic interpretation of WM was not a step toward conceptual clarification.

We proceed by first offering a reconstruction of relevant episodes from the first calls for a new quantum mechanics in 1924 and the award of the Nobel prize to Born thirty years later. This part of the story is the work of JH: §2.1 describes the invention of MM from its immediate prehistory to the crystallization of opinions about QM at the Solvay Conference of 1927, in which Born's probability interpretation played an essential part; §2.2 reviews the Nobel prizes awarded to the founders of MM and WM between 1929 and 1933 and concludes with EPR, Einstein's fertile challenge to quantum mechanics coauthored with Boris Podolsky and Nathan Rosen; and §2.3 interprets the award of the Nobel prize to Born in 1954 by a new Nobel committee, then a generation away from the battles of the founders, as a final confirmation of the consensus enjoyed by the Copenhagen interpretations. Here the various assessments made by the historical actors that eventuated in Born's prize serves as a barometer of the changing consensus of interpretations of QM.

---

[7] Wallace, *The Emergent Multiverse* (2014). Vaidman, "Many-Worlds Interpretation of Quantum Mechanics" (2019).
[8] Goldstein, "Bohmian Mechanics" (2021).
[9] Fuchs, "QBism, the Perimeter of Quantum Bayesianism" (2010).
[10] Healey, "Quantum Theory: a Pragmatist Approach" (2010).
[11] Dieks, "Quantum Reality, Perspectivalism and Covariance" (2019); "Wigner's Friend" (2022).
[12] Rovelli, "The Relational Interpretation" (2022). Laudisa, "Relational Quantum Mechanics" (2021).

Part 3, by CR, discusses more recent ideas on the interpretation of quantum mechanics and in their light reconsiders the status of MM: §3.1 introduces these ideas; §3.2 compares them with those guiding the early development of the theory. The historical role of WM, including its function in Born's first probabilistic formulations is reexamined: from this perspective, the early success of WM hampered understanding by its misleading emphasis on the role of quantum states. In §3.3 CR illustrates this thesis by fancying a plausible counterfactual historical evolution, where WM and its associated psi-ontology does not occur. In §4 we review these serious matters in a light-hearted dialogue.

## 2. The events

### 2.1 Before Helgoland

Physicists demobilizing after World War I were confronted with an elaborate version of Bohr's quantum atom of which they had had only a brief glimpse before the shooting started. The original had a single electron either tranquilly orbiting the nucleus of a hydrogen atom in an elliptical path (a "stationary state") or giving rise to a spectral line in a transition between states (a "quantum jump"). The postwar model, codified in Arnold Sommerfeld's *Atombau und Spektrallinien*, first published in 1919, was a hive of electrons traversing synchronized orbits with diverse speeds and directions. This "multiply-periodic" system provided a fulcrum for the lever with which Bohr expected to move his semi-classical model into a fully quantum one. He called this lever the "Correspondence Principle." It provided that at some limit results obtained by quantum theory should agree with results computed from classical physics. Bohr had demonstrated its working for the observed quantum frequencies $\nu(n, n - \tau)$ arising from a jump from the $n$th to the $(n - \tau)$th stationary state: for jumps for which $n \gg 1$, $\nu(n, n - \tau)$ becomes asymptotically equal to the overtone $\tau\omega(n)$ of the classical frequency $\omega(n)$ of the electron in the $n$th stationary state.[13]

In a further leap, Bohr saw a correspondence between the terms of the Fourier expansion of the electric moment of the orbiting electron and the intensity of light emitted: the square of the amplitude $C_\tau(n)$ associated with the overtone $\tau\omega(n)$ should measure, in the correspondence limit, the intensity of the spectral line $\nu(n, n - \tau)$. Bohr's student Hendrik Kramers had the job of calculating the amplitudes $C_\tau(n)$ for various classical orbits and comparing the results to the measured (quantum) intensities. The scheme had promising success in pinpointing transitions for which $C_\tau(n)$ vanishes for a particular value of $\tau$. Spectral lines that would correspond to such values of $\tau$ do not occur. These results earned Kramers his doctoral degree and a prominent place among quantum theorists.

A quantum quantity was now needed to relate to $|C_\tau(n)|^2$ as $\nu(n, n - \tau)$ did to $\tau\omega(n)$. To his great regret, Einstein provided the answer. In 1916 he had introduced "transition coefficients" to enable a derivation of Planck's radiation law, which he rated more fundamental than any of the several ways previously known. It was certainly the simplest. It describes the equilibrium reached in a collection of atoms capable of transitions between stationary states $m$ and $n$ when bathed in radiation of density $\rho(m, n)d\nu$ and frequency $\nu(m, n)$. Let an atom in state $m$ have, in unit time, probabilities $\rho(m, n)B(m, n)$ and $\rho(m, n)B(n, m)$ of emitting or absorbing a quantum $h\nu(m, n)$ under the stimulation of the radiation, and a probability $A(m, n)$ of spontaneously emitting such a quantum. In illustration of this last mode, Einstein mentioned radioactive decay, generally then regarded as a matter of pure chance, rooted in the nature of things. He may not have held this view, but his reference to it allowed an easy assimilation of spontaneous emission with radical probability.[14]

The energetic postwar generation of atomic physicists, which included Pauli, Heisenberg, and Jordan, soon ran into difficulties. Among the most serious was the discovery that they

---

could not calculate the spectra of higher elements from multiply periodic electron orbits with the precision Bohr had obtained in the one-electron case. They persuaded themselves that what they could not accomplish could not be accomplished and identified the difficulty with the concept of orbits; by early 1924, a year and a half before Heisenberg's breakthrough on Helgoland, a few physicists were calling for a calculus that invoked only observable quantities, precisely the dictum with which Heisenberg opened his paper on the "reinterpretation" of quantum theory. Sommerfeld explicitly abandoned electron orbits in treating complex spectra in the fourth edition of his *Atombau* (1924), for which he earned the praise of his former student Wolfgang Pauli. "I found it particularly nice that you set aside all model talk. Model concepts are now in a severe crisis of principle that I think will end only with a more radical sharpening of the difference between classical and quantum theory... The concept of definite, unambiguous orbits of electrons in the atom can hardly be maintained... In models we speak a language that does not do justice to the simplicity and beauty of the quantum world."[15]

Suspicion of multiply periodic orbits blunted physicists' prime tool for establishing correspondence between quantum and classical quantities. Bohr himself acknowledged this difficulty early in 1924 in an extraordinary scheme devised with Kramers following suggestions from an American visitor to Copenhagen, John Slater. The Bohr-Kramers-Slater (BKS) theory replaced the orbiting electrons with "virtual oscillators" vibrating at the quantum frequencies $\nu$. These oscillators created a virtual field that induced transitions, or, rather, probabilities for transitions, in atoms it reached.

BKS theory not only eschewed orbits and admitted radical probability, it sacrificed the canons of conservation of energy and momentum in individual atomic events. That is because the "radiation" from these virtual oscillators was as ghostly as they. To each atomic state $n$, BKS allotted a finite set of oscillators to account for down transitions $n \to (n - \tau)$ and an infinite number for up transitions $n \to (n + \tau)$. These virtual oscillators continuously emit their ghostly radiation at the frequencies $\nu(n, n - \tau)$ so that a collection of atoms resembled an orchestra of silently sounding pianos, silent because the virtual radiation conveyed no energy but only an inducement to change state. When a particular atom changed its state spontaneously with probability $A(m, n)$, it emitted light of frequency $\nu(m, n)$ and changed its virtual tune to the notes appropriate to its new state $n$. Meanwhile another atom in the collection in the state $n$ realized its probability $B(n, m)$ of absorbing an eligible quantum of light and rose from state $n$ to state $m$. The two acts had no direct connection. Not only were individual acts unpredictable, but conservation could be assured only statistically.[16]

---

[15] Pauli to Sommerfeld, 6 Dec 1924, in Pauli, *Wissenschaftlicher Briefwechsel*, **1**, 182.

Most physicists did not care to sacrifice conservation principles and rejoiced when experimenters destroyed BKS theory by demonstrating that in the Compton effect energy and momentum are conserved in individual events. Walter Bothe, who would share the Nobel Prize with Born, collaborated with Hans Geiger in the earliest and most decisive of these experiments; it is worth noting that Bothe understood energy density of radiation in Einstein's terms as the probability that a volume element contained one or more light quanta.[17] But the Bothe-Geiger experiment, though of immense importance in limiting flights of fancy, did not cancel the virtual oscillators or the conceptualization of quantum mechanics as a calculus of observables. Among those who clung to these elements of BKS were Kramers and Born. Kramers made good use of them in developing a dispersion theory in which virtual oscillators represented the unknown quantum mechanism of the scattering of light. In recommending his approach he stressed that it "contain[ed] only such quantities as allow of a direct physical interpretation on the basis of the fundamental postulates of the quantum theory of spectra and atomic constitution, and exhibit[ed] no further reminiscence of the mathematical theory of multiply periodic systems." He elaborated his ideas with the aid of the Correspondence Principle and in collaboration with Heisenberg, who thus thoroughly assimilated techniques and viewpoints he would need on Helgoland.[18]

Pauli had sketched something like the program Heisenberg would follow in a letter to Bohr written at the end of 1924, when experiment was burying BKS's "fake solution." The situation was dire, Pauli wrote, even menacing; the valence electron of alkali atoms was behaving duplicitously, with a "classically indescribable ambiguity," and so apparently was unable to afford leverage to the Correspondence Principle. "I have avoided using the concept of orbit in my [current] work." Somehow energy and momentum were more real than the motions to which they pertained. "The goal we are striving to attain must be to deduce these and all other physically real, observable characteristics of the stationary states from (integer) quantum numbers and quantum theoretical laws. But we must not shackle atoms to our prejudices (to which, in my opinion, the assumption of electron orbits belongs)."[19]

---

[16] BKS, in Van der Waerden, *Sources*, 166.
[17] Fick and Kant, *Stud. hist. phil. mod. physics*, **40** (2009), 398-9.
[18] Kramers, *Nature*, **114** (1924), 310-11 (dated 22 Jul 1924), in Van der Waerden, *Sources*, 201; Heisenberg and Kramers, in ibid., 223-51, esp. 225, 227, 232-4.

Pauli expressed these views in the context of his exclusion principle, which he had reached by assigning more quantum numbers than usual to atomic electrons to express nature's duplicity. Heisenberg greeted the principle with great jocularity. "You have caried the Schwindel [a cross between swindle and madness, a term of art among the Copenhagen brotherhood] to a hitherto unimagined height of craziness and beaten all previous records that you complained that I had set...by introducing electrons with four degrees of freedom." But be joyful, Heisenberg went on, Sommerfeld will welcome you to the land of formalism philistines, whose spectral numerology eschewed physical principles. Bohr answered that Pauli's addition to the Schwindel had put pressure on the Correspondence Principle but had not broken it. Perhaps the four quantum numbers distinguish between the inner and the outer space of the atom? Perhaps BKS is not dead. "I feel that we are approaching a decisive turning point since the extent of the entire Schwindel has been so fully characterized." Pauli doubted that the turning was nigh and set off for Sommerfeld's "Institute for Number Mysticism" to talk with Heisenberg.[20]

During 1924 Born also became skeptical about electron orbits and called for a calculus of observables to which he hoped to be guided by the Correspondence Principle. For example, he thought that orbiting electrons might perturb each other's motion in the same unmechanical way that light acted on them. He followed the lead of BKS in taking the virtual radiation from the electrons as a stimulus to transition and obtained Kramers' dispersion formula; but he declined explicitly to follow Kramers' identification of the $|C_\tau(n)|^2$ with Einstein's probability coefficients. This is a good example of his tendency to lose confidence in his judgment and to claim less than his due. Typically, he credited Bohr with clarifying his ideas and Heisenberg with helping with the calculations; critical readers, like the judges in Stockholm, might have wondered how much he contributed to papers in which he needed so much assistance. His generosity and unassertiveness may well have led his colleagues to underestimate him.[21]

[19] Pauli to Bohr, 12 Dec 1924, in Pauli, *Wissenschaftlicher Briefwechsel*, **1**, 187-9; "Scheinlösung," in Pauli to Sommerfeld, 6 Dec 1924, in ibid., 183-4.
[20] Heisenberg to Pauli, 15 Dec, Bohr to Pauli 22 Dec and reply, 31 Dec 1924, in Pauli. *Wissenschaftlicher Briefwechsel*, **1**, 192, 194-5, 198.

In lectures on *Atomdynamik* given in 1924, Born was bold enough to assimilate quantum jumps to radioactive decay, for which no cause could be assigned. He regarded the book in which he refined these lectures as the first of two volumes; the second, which he postponed to a distant future, would present the "final" quantum mechanics. As signposts to this future, he listed Heisenberg's account of complex spectra, a "deep and mysterious paper," he wrote Einstein, full of the current madness: duplicitous electrons, BKS theory, his own translations of the perturbation techniques of celestial mechanics into the embryonic speech of quanta, and the equivocal guidance of Bohr. In the preface to the English translation of *Atomdynamik*, published in 1927, Born reported that Heisenberg and Schrödinger had solved many of the problems he had identified in 1924. He gave as his reason for authorizing an English translation of his superannuated text the need he felt to make clear the struggles and genius of Bohr and the sources on which Heisenberg and Jordan drew. *Atomdynamik* "prepared the way for the central step which we owe to Heisenberg."[22]

Born had tried to take this step with Jordan's help. In a joint paper on the quantum theory of aperiodic processes, they announced that "the true laws of nature adopt only such quantities as are in principle observable, and determinable."[23] They sent this paper for publication in June 1925. A month later, Heisenberg handed Born the manuscript he had written after his productive sojourn in Helgoland. We know something of Born's frame of mind when he received it from a letter he sent Einstein three or four days later, on 15 July. He had been doing "very ordinary stuff," he wrote, very little in comparison with what an Einstein or a Bohr produced. Much of the time he was just puzzled. What was he to make of Louis de Broglie's wave-particles and the "rather mysterious differential calculus on which the quantum theory of atomic structure seems to be based"? He had turned to busy work. "Jordan and I are systematically…examining every imaginable correspondence relationship between classical, multiply-periodic systems and quantum atoms." Where would it lead? "I still feel very unsure of myself." How could any mere mortal keep up with Jordan or Heisenberg, "[whose] latest paper, soon to be published, appears rather mystifying but is certainly true and profound." In his later commentary on this letter, Born supposed that he had had in mind Heisenberg's Helgoland paper, but the dates make that impossible. The "mystifying" paper was Heisenberg's unravelling of the multiplet spectra using duplicitous electrons and quantum numerology from Sommerfeld's school.[24] When he wrote Einstein on 15 July, Born had not looked at Heisenberg's new manuscript.

When he reached its end he understood why its author had worried that its approach was "far too rough." It took off from the mid-summer madness over the intensities of the multiplet lines and managed, by removing references to electron orbits in calculating transition amplitudes, to lose the notion of steady state.[25] Much of Heisenberg's procedure, however, would have been familiar to Born. It included replacing the orbital quantities $xv$ making up angular momentum with the product of Fourier series and translating the quantization of angular momentum into an expression involving only observable frequencies $v(n, n - \alpha)$ and amplitudes $a(n, n - \alpha)$ using a trick that Kramers had employed in his dispersion theory. But there was something dramatically new in the mathematics: to translate the form for the product of two Fourier series, Heisenberg had made "the simplest and most natural assumption," which required summation (over $\alpha$) of forms like $a(n, n - \alpha)b(n - \alpha, n - \beta)$. As Heisenberg pointed out, in general the summation was not commutative.[26]

To do some physics, Heisenberg took over the classical formula applicable to the most difficult case he could solve, the anharmonic oscillator; expressed the formula in Fourier series; translated the series into quantum equations; and, invoking the quantum condition, obtained the frequencies and amplitudes as functions of $n, h$, and the parameters defining the oscillator. He gave no justification for using the classical equation of motion for the oscillator any more than Bohr's followers had done for using the Coulomb force in their atomic models. And he did not comment on the fact that his solution -- an array of numbers -- referred to the entire collection of possible spontaneous radiative transitions of an atom, rather than to its current state.

## 2.2 After Helgoland

Born soon recognized Heisenberg's novel sums as matrix algebra and turned the quantum condition into the inspiring form

$$\mathbf{pq} - \mathbf{qp} = \frac{h}{2\pi i}\mathbf{1}, \tag{1}$$

where the matrices $\mathbf{p}$ and $\mathbf{q}$ are avatars of the classical quantities' momentum and position and $\mathbf{1}$ is a matrix with all its diagonal elements equal to 1 and all other elements 0. Heisenberg rated this insight as "very shrewd."[27] Born was so impressed by it that he wanted the equation carved on his gravestone. There is profound symbolism hidden here, for the non-vanishing imaginary commutator contains the doctrine of radical probability. Born's gravestone thus stands as an everlasting challenge to Boltzmann's, which bears the legend $S = k \log W$. $W$ is the probability of ignorance arising from the incompetence of physicists unable to solve $10^{23}$ simultaneous equations of motion. As everyone knows, $S$ is entropy and $k$ Boltzmann's constant, which refers to a mystery no greater than the incredible smallness of atoms.

---

[26] Heisenberg, in van der Waerden, *Sources*, 265, 276 (quotes).
[27] Heisenberg to Pauli, 18 Sep 1925, in Pauli, *Wissenschaftlicher Briefwechsel*, **1**, 237.

Born had trouble proving that the off-diagonal elements of the left side of his matrix formula vanished. Heisenberg was not around to help; he was taking another vacation to prepare to work with Bohr in Copenhagen. Born tried Pauli. The enfant terrible answered that he would have no part in suffocating Heisenberg's physical ideas in Göttingen mathematics. So Born turned again to Jordan, who quickly found the needed proof.[28] Jordan contributed at least as much as Born to the subsequent development of the matrix formalism, some say the major part; but no safe determination of the relative importance of their contributions can be reached from the available evidence.[29]

Later in 1925 Heisenberg joined Born and Jordan for an elaborate extension of the theory, further aggravating, or rendering impossible, the apportionment of credit. When they were done the three had a complete formal theory that expressed dynamical relations via matrices $p$ and $q$ obeying the Hamilton-Jacobi equations and a way to obtain the matrix elements that measured the transition amplitudes. One needed only to find a $p$ and $q$ that simultaneously satisfied the Hamilton-Jacobi equation for the relevant Hamiltonian and the quantum condition, and also made the energy matrix diagonal. The method was too general for practical problems. Only after weeks of hard work did the disdainful Pauli succeed in diagonalizing the Hamiltonian of the hydrogen atom and so obtain its quantized energy levels.[30] Whether MM would have stultified without the unexpected aid of wave mechanics is of course unknowable, but we can say securely that physicists from all schools eagerly employed Schrödinger's methods in their calculations.[31]

What part did Born play in his collaboration with Jordan and Heisenberg? Jordan's recollections scarcely have a place for him. "It was a small number of very junior physicists who broke through the impasse under Werner Heisenberg's direction, our work was playfully called 'physics of 20-year olds'."[32] Nor did Jordan mention Born in discussing probability in quantum physics. Since this mistreatment comes from a book Jordan published and republished during World War II, it might be construed as a prudent concession to Nazi attacks on Jewish physics. But Jordan had defended the physics of Bohr and Einstein against this attack and after the war he again omitted Born from an account of quantum probability. And in late reminiscences about the beginnings of quantum mechanics, he observed that Heisenberg followed a well-articulated program to his breakthrough on Helgoland and singled out Kramers as a leader in it. Its architect, however, was Bohr. The deeper Jordan and Born dug into quantum mechanics, "the more vivid our understanding of Niels Bohr's guiding ideas [became]." Despite his enhanced understanding, Born did not pull his weight in founding quantum mechanics. According to Jordan, Born spent most of his time during their collaboration in a sanitorium, prohibited by his doctor from risking further mental strain after he had divined the principle (1).[33] He recovered enough to lecture on MM at the Massachusetts Institute of Technology in the winter of 1925/6. Again, he ended unassertively: we cannot tell yet whether we will be able to conceive of atomic processes in space and time; it is still possible to think in terms of orbits; "our theory...gives at most the probability of the jumps."[34]

Schrödinger was also hard at work that Christmastide and early in 1926 surprised the Göttingen group with WM. Contrary to MM, it employed mathematics familiar to physicists, offered a realistic model of stationary states, solved the hydrogen atom relatively easily, and owed nothing to the translation program of the Correspondence Principle. Soon Schrödinger and others showed that matrix elements could be computed from the solution $\Psi$ (the wave or psi function) of the fundamental equation of his theory. Planck, Einstein, and most other theorists were delighted, Bohr lukewarm, Heisenberg outraged at the "trash" (Mist) that challenged his breakthrough. "I found it horrible."[35] But there could be no doubt about the greater convenience, even necessity, of Schrödinger's methods of calculation, notably in collision problems of interest to Born.

---

[32] Jordan, *Physik* (1949), 15 (quote), and *Physik* (1945), 67-74.
[33] Jordan, *Physik. Blätter*, **31**:3 (1975), 98 (quote), 100.
[34] Born, *Problems* (1926), 128-9.

Born's interest in atomic collisions had been awakened by experiments performed by his Göttingen colleague James Franck and by an odd phenomenon uncovered by Carl Ramsauer, then at Heidelberg. Franck's experiments inverted the Nobel-prize-winning work he and Gustav Hertz had performed before the war. Then they produced quantum jumps in unexcited atoms by bombarding them with electrons; now he reduced an excited atom to its ground state via a colliding atom that carried away the released quantum as kinetic energy. Franck's "collisions of the second kind" were easier to conceive than Ramsauer's discovery that slow electrons penetrate further through noble gases than fast ones. To these intriguing puzzles, Rutherford's Cambridge laboratory added the challenge of deviations from the scattering results that had prompted his invention of the nuclear atom.[36]

Born attacked these problems using his translation rules and the Correspondence Principle and published his modest progress in the spring of 1925 in the paper on aperiodic processes he wrote with Jordan. A year later, finding no way to calculate the scattering of an electron by a gas molecule using MM, he supposed that before and after the collision the atom rests in a stationary state and that the separated electron can be described as a plane de Broglie wave. The struck atom can change state by a collision of the first or second kind while the electron, conserving energy, enters asymptotically into some "definite, rectilinear, constant motion." Definite but unpredictable. WM allows the calculation only of the "diffracted" outgoing wave function $\Psi$ and through it the relative probabilities of the final states. "If a corpuscular interpretation [of them] is wanted, there is only one interpretation possible." $\Psi$ defines the probability that the incoming definite motion is "thrown" into a particular final one.[37]

Born thus returned to MM's concept of transition probabilities from which he had separated himself at the beginning of his analysis. "I intentionally avoid the term Übergangswahrscheinlichkeiten," he wrote, and indeed it is a mouthful. Instead he would temporarily leave open the nature of transitions and investigate what quantum mechanics, if a logically complete and closed system, said about them. That recommended pursuing his work on collision processes, in which, according to Bohr, all the main difficulties of quantum physics turned up. Only Schrödinger's method gave access to the scattering domain and, "precisely for that reason, [Born] regard[ed] it as the deepest foundation of quantum laws". And he was right, it did throw light, although not of full day, on the nature of jumps. "No answer is given to the question, 'what is the state after the collision,' but only to the question, 'how probable is a given state of the collision'?"[38] And so Born came back to transition probabilities, but with a significant extrapolation. The probability that the electron scatters into a given state became the probability that it would be found in that state or, more problematically, as more readily assimilated to psi-ontology, that it occupied that state.

Should we hope someday to discover ways to determine in advance the state of the scattered particle? "Or should we believe that the agreement of theory and experiment in their inability to give conditions for causal behavior is a pre-established harmony arising from the non-existence of such conditions? I am inclined to give up determinateness [*Determiniertheit*] in the atomic world. But that is a philosophical problem, for which physical arguments alone are not decisive."[39] This first presentation of the probabilistic interpretation of the wave function does not suggest strong conviction, although its labored form and weak conclusion may have owed something to Born's habitual caution and self-doubt. Heisenberg had trouble untangling its meaning; Pauli had to inform him that it interpreted $|\Psi|^2 dq$ as the likelihood of finding a particle within a volume $dq$.[40]

In July 1926, a month after declaring his "inclin[ation] to give up determinateness," Born explained his purpose in a way that may have damaged his claim to originality. He saw himself as pursuing a middle way, between the statistical approach built into MM and the realistic interpretation Schrödinger sought to give his $\Psi$ wave; and he designated Einstein's concept of electromagnetic waves as a "ghost field" guiding photons as the inspiration for his third way. Like the soundless symphonies of the BKS theory, which Born had favored, Einstein's ghost field determined a probability, in this case of finding a photon at a given place and time. Born's idea was to interpret $\Psi$ as the ghost field for de Broglie's electron wave.[41] In a lecture given at the physicists' international jamboree on Lake Como in 1927, the same meeting in which Bohr first mumbled publicly about complementarity, Born identified his contribution to the statistical account of quantum processes as an improvement on BKS. It was important: his theory conserved energy and momentum as required by the formalism of MM and the experiments of Bothe and Geiger.[42]

"Guide" is not quite the right word for the relation between these ghost fields and their particles; "advisor" would be better; for although $\Psi$ is in full command of itself (it develops in time like a classical quantity), its influence on the electron is only advisory. Born arrived at the "paradox" that "the motion of the particle follows probability laws while the probability itself [$\Psi(t)$] spreads in agreement with the law of causality." Once again, however, he did not insist. He observed that anyone dissatisfied with his probabilistic treatment was free to assume the presence of unspecified parameters that determine individual events. They would not bring in new results, however, since taking them into account would necessarily lead to the same formula for confirmed experimental results. In short, Born deemed a non-probabilistic account of quantum mechanics "improbable."[43] In this permissiveness, some commentators have discovered the origin – or license – for hidden variable theories.[44] Others have found a vagueness whose needed clarification was an important spur to further development.[45]

Born continued to permit belief in "micro-coordinates" at least through March 1927.[46] Six months later he expressed himself more boldly. The occasion was a report on the then current state of quantum mechanics to the fifth Solvay conference held in Brussels in October 1927, where Einstein first did formal battle with the new theories. Born's boldness may be associated with the confidence of his co-rapporteur, Heisenberg, who probably wrote most of their joint paper; in any case he was almost certainly the author of the parts where poor editing left "I" where "we" belonged. The statistical element in the theory was now irremediable, arising from "an essential impotence, profoundly rooted in in the nature of our ability to understand physical phenomena." Indeed, it was there all along in MM and needed only to be demonstrated for WM. "The best confirmation of this point of view is found in collision phenomena." Born and Heisenberg credit this confirmation to Born and to Einstein. Bohr's talk at Solvay referred to Born's contribution in a similar vein: "he succeeded in giving a statistical interpretation of the wave function...required by the postulate of quanta."[47] Good work, but obvious.

The Solvay meeting did not end in full agreement about the meaning of quantum mechanics. Owing to the argumentative presence of Bohr, Kramers, Pauli, Born, Dirac, and Heisenberg, however, a consensus emerged in favor of the Copenhagen-Göttingen position, which by then had incorporated the view that quantum mechanics in all its forms embraced radical probability and could give only a statistical prognosis of the outcome of individual atomic events. Ever optimistic, Heisenberg thought that the discussions had silenced the opposition. He described the triumph to his parents: "Bohr's and my views were generally accepted, and in any case there are no more serious objections, even by Einstein and Schrödinger."[48] What Heisenberg soon christened the Copenhagen spirit had begun its hegemonic flight. Born's courage rose with it. A year after the Solvay meeting, having recovered from a nervous breakdown he ascribed to his efforts to keep up with Heisenberg and Jordan, he informed the Göttingen Academy of Sciences that if it were possible to circumvent radical probabilities in individual events, quantum mechanics would be wrong, a possibility he thought extremely unlikely.[49]

---

[46] Born, *Naturw.*, 15 (11 Mar 1927), 241, = Born, *Physik* (1966), 12. Cf. Im, *Arch. hist. exact sci.*, **44** (1996), 97-9.

[47] Born and Heisenberg, in *Electrons et photons* (1928), 144, 160 (1st quote), 165 (2nd); Bohr, in ibid., 233. Bacciagaluppi and Valentini, *Quantum theory* (2009), 372-401, give a translation of the paper, and, on pp. 80-107, a useful commentary on it.

[48] Heisenberg, *Liebe Eltern* (2003), 126.

Jordan endorsed Born's interpretation in a lecture given in Göttingen in 1927, translated into English by Robert Oppenheimer, and published in *Nature* to the Anglophone world. De Broglie conceded the necessity of Born's views without mentioning Born in a widely read journal dealing with metaphysics and ethics; Heisenberg did the same, spreading the good tidings among general audiences of German academics.[50] Then came the founders' textbooks, several of them, in 1929 and 1930: Pauli's *Allgemeine Grundlagen der Quantentheorie des Atombaues*, De Broglie's *Introduction à l'étude de la mécanique ondulatoire*, Heisenberg's *Physical principles of the quantum theory*, Born and Jordan's *Elementare Quantenmechanik*, Dirac's *Quantum Mechanics*. None of these texts bothered much about Born. Even Born and Jordan mentioned his papers of 1926 only once, in a footnote that also cites papers by BKS, Bohr, Heisenberg, Dirac, Jordan, Heisenberg, von Neumann, and Pauli.

In his text, Pauli observed that a statistical interpretation of $\Psi$ was suggested by its location in phase space, rather than in ordinary space, and by old analogies between waves and particles, and credited Born with emphasizing these inferences in treating collision processes. Heisenberg attributed to Born the idea that $\Psi(x, t)\Psi^*(x, t)$ represents the probability of finding an electron at $x$ and $t$; but as here Heisenberg had in mind stationary states, not collisions, he warned (with scare quotes) against using the "'statistical interpretation'...too schematically." De Broglie introduced Born's probability as something obvious "after a little reflection" and assigned Born the merit of having "foreseen" the need to abandon determinism. Taciturn Dirac permitted himself to declare that, in quantum mechanics, "The most that can be predicted is the probability of occurrence of each of the possible results."[51]

During the 1930s a standard interpretation of QM settled around complementarity. Its main ingredients were Bohr's doctrine developed in a positivistic direction, Born's understanding of probability, Heisenberg's uncertainty principle, and, as an associated resource, a reduction of the theory to a general mathematical form by von Neumann. Many physicists believed without troubling to study von Neumann's treatise of 1932 that he had accomplished the impossible and proved that no theory of hidden variables could be compatible with QM.

A few psi-ontologists kept up the fight, however, which came to a head in 1935 with two efforts to discredit the standard theory. One employed ridicule. The standard probability reading applied to a macroscopic object seemed to imply the existence of such oddities as Schrödinger's caged cat, which was both dead and alive, or was neither dead nor alive, until a quantum physicist looked at it. The other, the work of Einstein and his junior collaborators Podolsky and Rosen, charged QM with incompleteness. They argued that if two particles $A$ and $B$ interact and then drift so far apart that they can no longer influence one another, examination of one can yield more information about the other than QM allowed. An initial permissible experiment can give simultaneous measurements of the particles' separation $q_A - q_B$ and momentum sum $p_A + p_B$. Subsequently, an experimenter operating on A can decide whether to obtain an exact value of $q_A$ or $p_A$. If $q_A$, then $q_B$ is known via the measured initial separation; if $p_A$, then $p_B$ is known from the initial momentum sum. Since exact values of both $p_B$ and $q_B$ can be inferred without molesting $B$, EPR concluded that $B$ has a $p$ and a $q$ with such values, in contradiction to QM.[52]

In the ensuing discussion, Bohr showed that a detailed analysis of the possibilities of measurement nullified EPR's argument: the experimental arrangements needed for an exact $p$ measurement are incompatible with those needed for an exact $q$ measurement. The experimenter makes the choice for $A$ and $B$ somehow follows suit; if the choice is $q_A$ (or $p_A$), the possibility of measuring an exact $p_B$ (or $q_B$) disappears. In other words (the words are Schrödinger's), $A$ and $B$ "are entangled;" even at a great distance they retain some vestige of their initial interaction. Most physicists accepted Bohr's answer as definitive and the attack of EPR only strengthened the hold of the standard interpretation.

---

[51] Born and Jordan, *Quantenmechanik* (1930), 288; Pauli, *Grundlagen* (1929), from Müller-Pouillet, *Lehrbuch*, **2**:2, in Pauli, *Coll. sci. papers*, **1**, 763; Heisenberg, *Principles* (1930), 33-4; De Broglie, *Introduction* (1930), quoted from the English edition, also of 1930, pp. 5, 9; Dirac, *Quantum mechanics* (1930), 4.
[52] The exchange took place in the *Physical review* for 1935; Jammer, *Philosophy* (1974), gives a plentiful account of the arguments and responses to them.

It did not reconcile Einstein to Born's elucidation of $\psi$ waves. "I still do not believe that the statistical methods of quantum theory are the last word," he wrote Born, "but for the time being I am alone in my opinion." To this remark Born later made the revealing comment that, although complementarity held the statistical interpretation as a core concept, "I consider it unjustified that this is usually cited as having originated in Copenhagen."[53] Let us acknowledge that the philosophy of science "culminates in Niels Bohr's philosophy of complementarity;" but we need not suppose that every significant ingredient in it originated with him.[54] Was the probabilistic interpretation prize worthy? When applied to scattering, was it not, as Born wrote later, self-evident?[55]

## 2.3 Born's Prize

Knowledge of a few rules of the Nobel process is necessary to appreciate the potholes in Born's long road to the Nobel Prize. An award began with nominations by people who possessed permanent rights to make them (previous prize winners, members of the Royal Swedish Academy of Sciences (KVA), professors of physics at Scandinavian universities, and members of the Nobel Committee for Physics of KVA) and individuals and members of physics departments specifically invited to propose candidates for a given year. On average, 37 of the pool of eligible nominators exercised their franchise each year during the 1930s.[56] Nominations had to be received before 1 February to be considered for the year's prize; late

---

ones were applied to the following year. Many complicated alternatives and divisions of prizes were proposed and considered subject to the rule that a given prize would not be shared by more than three people. The nominations went to KVA's Nobel Committee for Physics, which commissioned reports on candidates it considered prizeworthy. The reports played an important part in the committee's deliberations before it recommended a candidate to the Academy. If the committee could not agree on any candidate, it reserved the prize for the following year; but if it failed two times running, the Academy canceled the prize and retained the money, a portion of which went to the committee. The subsequent award of a reserved prize created two winners, or sets of winners, at the same ceremony. For example, in 1921 the committee could not find a worthy laureate, although neither Einstein nor Bohr had yet been honored. In 1922 both received it, Einstein's being the reserved prize.

By 1930 the physics establishment had realized that the invention of quantum mechanics deserved a Nobel Prize. Whom to award, however, presented unusual difficulties: the number of competitors was small enough to recognize all, but all did not deserve equal credit. The case was easier for WM because its creators were loners, Louis de Broglie and Erwin Schrödinger; but MM, which came first, was the product of a program set forth by Niels Bohr and developed by Hendrik Kramers, Max Born, Wolfgang Pauli, Werner Heisenberg, and Pascual Jordan. And these two approaches were not the only ones; another loner, Paul Dirac, inspired by early notice of MM, created a powerful idiosyncratic version that had much in common with some ideas of Jordan. A good indication of the challenges of assigning credit appears from a complicated proposal drawn up by Einstein in September 1928 for the prize of 1929.

Einstein began his proposal with conviction: "the most important, certain, and unawarded work in physics is the knowledge of the wave nature of material processes." Therefore, the prize should go half to de Broglie and half to the American experimentalists, whoever they were, who demonstrated it. But then there was the difficulty that de Broglie did not fully work out the idea he introduced. Perhaps then Heisenberg and Schrödinger? They were certainly worthy, but on second thought Einstein came back to de Broglie whose ideas had been tested, whereas the grandiose theories of the others might not survive. Even with de Broglie, however, the business was not clear; not only had he not taken the decisive step alone, but it was also not obvious (on third thought) that the other half should go to those Americans. Maybe a theorist should share; no one would object to de Broglie and Schrödinger. Another possibility of the same kind was Heisenberg-Born-Jordan. "But Born and Jordan have done less than Heisenberg and this solution would consider only theoretical contributions, which in this case certainly are the greater, but not the most secure results. I'm happy not to have to make the choice myself." In the end, Einstein decided that choosing prize winners was more difficult than doing physics and nominated no one.

There is no evidence that Einstein ever again considered recommending his friend Born to Stockholm. Instead, he stuck to an evaluation he had formed long before the invention of MM. When asked in 1912 for an opinion of Born's work, Einstein allowed that he was "a good calculator, but so far has not shown much acuteness about physics." It might appear that by recommending Born as "the most important (with [Pieter] Debye) of theoretical physicists" in 1920, Einstein had changed his mind.[57] He then thought that Born had become more concerned with facts and that his work on crystals showed promise; but that, nonetheless, his strength remained in mathematics and a capacity for hard work. Einstein's enduring estimate appears in his response to a letter from Paul Ehrenfest. To Ehrenfest's remark that, despite his relative nothingness, Einstein and Bohr had always supported his work, "whereas contact with other theorists totally discourages me," Einstein replied that like himself and Bohr, Ehrenfest was a *Principienfuchser*, a worrier about foundations, while most other theorists were virtuosi, polished mathematicians or devotees of detail, but not quite the real thing. He gave as examples of the polished virtuoso Born and his predecessor at Göttingen, Debye.[58]

Einstein's first notion of a proper division of the spoils agreed with the thinking in Stockholm. The prize of 1929 went undivided to de Broglie, who received a dozen nominations against 10 for Schrödinger, 9 for Arnold Sommerfeld, and 3 for Heisenberg. The prize for 1930 went to C.V. Raman for experimental work in quantum optics on the recommendations of Bohr, de

---

[57] Einstein to Alfred Kleiner, 3 Apr 1912, and to David Hilbert, 21 Feb 1920, in Einstein, *Collected Papers*, **5**, 445, and **9**, 440; cf. Greenspan, *End* (2005), 62, 96.
[58] Ehrenfest to Einstein, 16 Sep 1925, and reply, 18 Sept, in Einstein, *Collected Papers*, **15**, 123, 126. The concept of *Prinzipienfuchser* is developed in Baggott and Heilbron, *Quantum drama* (2024).

Broglie, Rutherford and seven others.[59] The Swedish pundits saw no clear and compelling reason to reward anyone in physics for 1931 or 1932. The honors for quantum mechanics were completed the following year, when Schrödinger and Dirac shared the prize for 1933, Heisenberg received the reserved prize for 1932, and the KVA kept the kroner for the prize of 1931.

In his Nobel lecture, Heisenberg mentioned Born's probabilistic interpretation "as a special case of more general laws and as a natural consequence of the basic principles of quantum mechanics," an ungenerous assessment he moderated by suggesting that future progress should be based on the work of de Broglie, Born, Jordan, and Dirac. He was not comfortable with this formulation either. He wrote to Born soon after picking up his prize that his designation without his collaborators "depresses me and I hardly know what to write to you... All good physicists know how great was your and Jordan's contribution to the structure of quantum mechanics–and this remains unchanged by a wrong decision from outside...[I] feel a little ashamed."[60] But only a little; Heisenberg never exercised his Nobel franchise on behalf of Born. Nor did Schrödinger, although he and Born were close friends, close enough for Schrödinger to berate "Maxel" for "the impudence with which you assert time and again that the Copenhagen Interpretation is practically universally accepted."[61]

The Nobel Committee would have done well to recognize Born and Jordan in 1933, perhaps by awarding half of the prize to Heisenberg and dividing the other between his collaborators. The prize Born ultimately received, for a tentatively proposed and obscurely phrased contribution, well within easy extrapolation of ideas held in Copenhagen and Göttingen when formulated, required the Nobel institution to shut its eyes to the historical context to right a historical injustice. That Born's probabilistic interpretation of the wave function fit perfectly with statistical descriptions of quantum transitions routine by 1926 accounts for its quick adoption by the Copenhagen and Göttingen schools. "We were so accustomed to

making statistical considerations, and to shift it one layer deeper seemed to us not so very important."[62]

Why not include him in Heisenberg's prize? One reason was formal: no one nominated either Born or Jordan for the prize of 1933. The antiquated reward system still sought the individual investigator, the lone inventor like Nobel, single-handedly breaking the barriers of received thought. The ideal did not correspond to the circumstances in which MM matured. Born later criticized himself for sharing this romance. "I gave so much prominence to Heisenberg that my own contribution to quantum mechanics received very little attention until quite recently."[63] He had learned his lesson. In his Nobel lecture, he immodestly claimed the "discovery of...the foundations of a new way of thinking about the phenomena of nature."[64]

Born received three nominations before 1945.[65] The first two were from experimentalists, Peter Pringsheim, who recommended that Born share with Heisenberg, and Dirk Coster, who proposed an undivided prize for Born's work on crystal structure and MM. The physics committee took these nominations seriously enough to order an evaluation by one of its members, Carl Oseen, a theorist close to Bohr with a serious interest in the history of physics. Oseen reported that Born's crystal work was incomplete, overly mathematical, and inadequately confirmed by experiment; no prize could be given for "such a half-right, over-elaborated [*övervässende*], deductive theory." Oseen thought that Born's best claim on a prize was his probabilistic interpretation of Schrödinger's wave function, although it suffered from the defect of "raising the whole problem of determinism." Setting that aside, how could Born be chosen without Jordan or given precedence over the brilliant but still unrewarded Pauli? The puzzle had no easy solution. Oseen observed that since no one had nominated Jordan, it would be advisable for the committee to attend to other fields. That proved too hard; it failed to find a candidate for the prize for 1934.[66] The awards for the remaining years before World War II went to deserving experimentalists. The third nomination for Born, made in 1939 by the Italian mathematician Tullio Levi-Civita, to honor his oeuvre in relativity, crystal structure, MM, the probability interpretation, and electron theory, brought Oseen back to the drawing board. He saw no reason to change his mind.[67]

No Nobel Prize in physics or chemistry was given during the first years of the second world war. In 1944, with the winner of the war now obvious, the Academy could be bolder. It gave the reserved prize for the previous year and the current prize to two American Jews, one naturalized after fleeing the Nazis, Otto Stern, and the other native born, Isidor Isaac Rabi, for their experimental work on magnetic properties of elementary particles (Stern) and nuclei (Rabi). On nomination by Einstein and others, Pauli received an undivided prize for the exclusion principle in 1945, the first given a theorist since 1933. Born's prospects brightened; Pauli's prize, for a discovery he had made 20 years earlier, could be considered an act of retrospective justice. For 1946, de Broglie proposed Born for his oeuvre in general. Two other mandarins and prize winners, Born's close colleague James Franck and Enrico Fermi, nominated him for 1947 for his crystal theory (although Fermi's letter arrived too late for consideration) and a professor at the University of Cracow, Czeslav Bielobrzeski, nominated him for his statistical interpretation, "[which] has revolutionized our conception of the laws of nature by revealing an indeterminism sui generis of atomic processes."[68]

Bielobrzeski knew his subject. He had organized a conference in Warsaw in 1938 on "New Theories in Physics" at which Bohr and de Broglie represented the founders and John von Neumann and Eugene Wigner the clarifiers of quantum mechanics. Bohr had begun his lengthy and important lecture, "The causality problem in atomic physics," with remarks on the introduction of statistical methods into quantum theory, which he credited to Born -- but also to Dirac and Jordan.[69]

---

[66] Letters from Pringsheim to KVA, 25 Jan 1930, and Coster, 14 December 1933; Oseen, "Utredning," in KVA, "Protokoll," 1934, Bil. 2, 46-7 (quotes), 49 (determinism), 54-5.

[67] Levi-Civita to KVA, 30 Dec 1938; Oseen, "Protokoll," 1939, Bil. 1, 15, 29.

[68] Letters to KVA from de Broglie, 18 Jan 1945 (received 9 Feb), and from Fermi, 30 Jan (received 7 Feb), Franck, 23 Jan, and Bielobrzeski, 16 Jan 1947.

For the prize of 1948 Born had four proposers, Fermi (the late nomination for 1947), Franck, and two Englishmen, all of whom emphasized crystal physics.[70] Franck repeated his suggestion in 1949 while two professors from the University of Cracow followed Bielobrzeski in emphasizing Born's statistical interpretation of quantum mechanics, "the decisive step from classical to modern physics." They were joined in this opinion by Erwin Madelung, although he was a crystallographer. Madelung hit on the recipe the Academy would later adopt by suggesting that the experimentalist Walter Bothe share the prize for work as far out of date as Born's.[71] Agitation for Born gained momentum in 1950, when Bielobrzeski repeated his proposal for 1947 and the big gun from Copenhagen broke its silence. Bohr and his colleague Christian Møller, professor of theoretical physics at the University of Copenhagen, made the singular but appropriate proposal that the prize be divided between Born and Kramers. They made an excellent complementary pair, Kramers for preparing the way for Heisenberg and Born for developing Heisenberg's ideas into MM. Another member of Bohr's circle, Torsten Gustafson, professor of physics at Lund University, seconded the proposed division and observed that, since the Academy had given the prize to Pauli for a stale contribution, it should have no difficulty in honoring Born and Kramers.[72]

Bohr took a deeper view of retrospective justice. It is important, he wrote, that at a time "when atomic theory has come to a new phase of development that everyone who works in the field should have the opportunity, again and again, to recognize the value of fundamental contributions to the foundations on which we must build." Consequently, the award he proposed would be widely applauded. One who would have been pleased, or perhaps half-pleased, was Heisenberg. He had nominated Kramers for 1949 primarily for his dispersion theory of 1924, "without which the later quantum mechanics would have been impossible;"

---

after all, it was the foundation on which he, Heisenberg, had built. Bohr and Møller repeated their proposals in 1951, but too late to be taken into consideration that year.[73] The proposals were carried over to 1952, but again could not be considered, because Kramers died before a decision could be made.

Born received a total of 16 nominations for the years 1951 to 1954. Franck and Fermi kept up their campaign. Several others mentioned Born's crystal theories, but most vied for superlatives to describe his contributions to quantum mechanics. The statistical interpretation, they said, forced "profound alterations in our Natural Philosophy... comparable only to the change of concepts due to the theory of relativity" (Walter Heitler, Zurich); it was "one of the keystones of contemporary physics" (Simone Franchetti, Florence), "one of the most fundamental concepts of modern quantum mechanics" (Hans Staub, Zurich), "basic for most later work" (Herbert Fröhlich, Liverpool), "a concept which has dominated all the development of theoretical physics since he formulated it" (Emilio Segrè, Berkeley).[74] Thus pressed, the Academy's physics committee asked its member Ivar Waller to evaluate Born once again.

Waller had undertaken the task before, in 1948, and had seen no reason then to disagree with Oseen. Born's work was not quite prizeworthy; the value of his statistical interpretation was uncertain (*tviksam*); his crystal theory, which Born himself admitted needed revision, was not as important as Franck and Fermi affirmed; and his other ideas had produced only interesting formalisms.[75] Six years later, Waller discovered that Oseen's objections to awarding Born a prize for the statistical interpretation had lost their force. *Nihil obstat*! Born could be honored without including Jordan and Pauli; Pauli had his prize and the statistical interpretation, as distinct from MM, belonged to Born alone. That was lucky. A shared prize with Jordan was scarcely possible; Born had fled the Nazis and Jordan had joined them. Moreover, Waller argued, because it had become a standard ingredient in quantum mechanics, the statistical interpretation gained in importance as quantum mechanics conquered new territories. The committee accepted Waller's revisionism and pointed to the very few nominations Born had received before 1945 as a reason for his belated recognition; but his merits did not seem great enough to deserve an undivided prize and the committee advised sharing it evenly between Born and Bothe.[76]

---

[73] Letters from Heisenberg, 20 April 1948, and from Bohr and Møller, 31 Jan 1951 (received after Feb 1).
[74] Letters from Heitler, 12 Dec 1950, Franchetti, 25 Jan 1952, Staub, 23 Jan 1953, Fröhlich, 26 Jan 1954, and Segrè, 28 Dec 1953.
[75] Waller, KVA, "Protokoll," 1948, 101-2.

Born was ecstatic. Although "conscious of [Heisenberg's] superiority," he had been irritated and depressed that he had not shared the prize for MM. But a prize had come at last, and for something he had done on his own, without the assistance of his brilliant students. Born ascribed the delay of three decades to opposition to his ideas by Planck, de Broglie, Schrödinger, and Einstein, and the eventual award to recognition that "my ideas had become the common property of all physicists." In this he thought that the Copenhagen school deserved most of the credit for endorsing "the line of thinking I originated." These remarks annotate a letter from Einstein congratulating Born on receiving his prize, "although strangely belatedly, for your fundamental contributions to the present quantum theory...Your subsequent statistical interpretation of the description...decisively clarified our thinking." And then, as if to deflect attention from his part in the strange belatedness, Einstein observed that the prize money would come in very handy because Born had just retired from his professorship.[77]

This story may explain well enough why Born was passed over in the awards of the early 1930s. It does not answer the question why he received it at all. No laureate nominated Born before the postwar wave in his favor began in 1946. Then came a short parade: Franck and Fermi (1947-49, 1954), and Bohr (1950, 1952), all of whom except Fermi (prize of 1938) had had a right of nomination since the 1920s. In contrast, Einstein, Heisenberg, Dirac, Schrödinger, and Pauli never nominated Born. And Bohr's very belated nominations of 1950 and 1952, although no doubt intended to correct an oversight, were to do justice more to Kramers than to Born.

---

[76] Waller, KVA, "Protokoll," 1954, 126-7, 131, and Nobel Committee for Physics, ibid., 12-15.
[77] Einstein to Born, n.d., and Born's later comments, in Born, *Born-Einstein letters* (1971), 228-9; Born, *My life and my views* (1968), 36-7.

An answer to the question is that by 1950 the founders were feeling their mortality. Kramers was the first to go; his illness may have prompted Bohr's suggestion of a prize shared by him and Born. The unbeliever, Einstein, died in 1955, Pauli in 1958, Schrödinger in 1961, Bohr in 1962. Bohr had become increasingly concerned to tell the story of "the great adventure" he had led; posterity needed the facts to appreciate it properly.[78] At his invitation a group of American historians came to Copenhagen to study the antecedents of quantum mechanics for a year. He endured them for a month or two before he suffered the stroke that killed him.

The award of the Nobel prize to Born in 1954 made a perfect emblem of the "great adventure." It brought out the main distinguishing feature of the new physics, its radically probabilistic basis, and implied its coherence and completeness by rewarding the co-inventor of MM for his elucidation of its rival WM. The earlier reservations about Born's contributions were ignored or forgotten. And so the Nobel establishment gave a fifth prize for wave mechanics, including those awarded in 1937 for the "experimental discovery of the diffraction of electrons." MM could boast but one.

With the passing of the founders a new breed of *Prinzipienfuchser* came sniffing, inspired by David Bohm's demonstration in 1952 that QM did not rule out hidden variables and kept optimistic by John Bell's later discovery that experiment might decide the matter.

## 3. An Alternative

There is more than one way to write a chapter of the history of science. An alternative to the standard method of reconstructing and evaluating events as they looked to contemporaries is to trace lines of development of what *later* became clear and important. From this second perspective, our evaluation of past shifts in our understanding of the significance of ideas and the impact of their legacy evolves. These alternatives are not entirely separable. The historian who follows the standard method cannot avoid being influenced by his or her situation in time and place. The commonplace that each generation writes history anew applies to the reconstruction of past science as well as to revaluations of political developments, although perhaps with less fanfare. The following revaluation of MM and Born's contributions to it, which draws on current versions of quantum mechanics, is not so different in its objectives from those of less explicit and more nuanced histories as to incur the stigma of whiggism. That offense, properly abhorred but often committed by good historians, treats history as a series of steps toward an ineluctable present necessarily wiser than the past. Opinions based on the present situation should not be taken as the final answer. Since the meaning of the quantum revolution is a subject of vigorous and healthy dispute, evaluation of the work of the founders should also be in flux.

---

[78] Bohr to Heisenberg, Nov and Dec 1961 and Mar 1962, probably unsent, in Dörries, *Frayn's Copenhagen* (2005), 119, 125, 135, 149, 169.

### 3.1 Current assessments

Quantum mechanics was born twice: first in 1925 as a matrix, or q-variable, mechanics, in the work of the Göttingen group[79] and Dirac's papers[80], and then, half a year later, as WM in the series by Erwin Schrödinger[81]. The two deliveries stemmed from markedly different conceptions.

Something of the original split persists in the current discussion. Interpretations such as Many Worlds or Bohmian mechanics have built upon Schrödinger's early intuitions that quantum physics is essentially about a continuous evolution of a wave function, or, more generally, of an objective quantum state $\psi$. In these interpretations, the mathematical object $\psi$ is assumed to represent an actual real entity existing in nature. In the jargon, $\psi$ is said to have an "ontological" interpretation.

In recent years, an opposite understanding of the theory has been growing. This includes relational, perspectival, and pragmatist interpretations, and QBism, among others. In these interpretations, the theory's mathematical quantities that have a direct physical interpretation are the "measurement outcomes" and their transition amplitudes. Measurement outcomes are represented by eigenvalues of physical variables. These, rather than $\psi$, are assumed to represent real facts of nature. The notion of measurement outcome can be generalized to any interaction between any two systems, and physical facts can be understood as relations between any two physical systems[82]. All these interpretations deemphasize the ontological role of the quantum state: $\psi$ is interpreted as a calculation tool, a bookkeeping device for the information one system can hold about outcomes of past or future interactions with another system. In particular, $\psi$ does not characterize a single system: it pertains to the relations between two physical systems.

---

[79] Heisenberg, "Über quantentheoretische Umdeutung kinematischer und mechanischer Beziehungen" (1925). Born and Jordan. "Zur Quantentheorie aperiodischer Vorgänge" (1925). Born, Jordan and Heisenberg, "Zur Quantenmechanik II" (1926). Pauli, "Über das Wasserstoffspektrum vom Standpunkt der neuen Quantenmechanik" (1926).
[80] Dirac, "The fundamental equations of quantum mechanics" (1925).
[81] Schrödinger, "Quantisierung als Eigenwertproblem (Erste Mitteilung)" (1926); "Quantisierung als Eigenwertproblem (Zweite Mitteilung)" (1926).

In the light of these interpretations, the core features of nature discovered in the 1920s with the formalization of quantum theory are those that guided Max Born and inspired the group of young students around him, not those hypothesized by Schrödinger in his early papers. Although Schrödinger's early notion that particles could be interpreted as wave packets cannot be maintained (and no modern theorist maintains it), it remains a misleading feature in many introductory textbooks.

A common path followed by introductory courses in quantum theory is to start from WM: present the Schrödinger equation upfront —perhaps with some historical or experimental motivation— and school students into computing various wave configurations in the presence of potential wells and similar artificial situations. The time-dependent Schrödinger equation appears as the successor of Newton's law and the textbook triumphantly produces the spectrum of the hydrogen atom from its time-independent version.[83]

This way of introducing quantum physics is far from commanding unanimity (it has been chastised as "a toxic habit that betrays history, logic, and reasonableness"[84]). It perpetuates the idea that quantum physics is about a "quantum state" $\psi$, an objective state evolving in time. An immediate consequence of this view is the brutality of the resulting measurement problem: if an electron is described by a wave literally diffused in space, how does it contrive to leave a point-like mark when it hits a screen?

_______________________

[82] Rovelli, "Relational Quantum Mechanics" (1996).

[83] See for instance Griffiths and Schroeter, *Introduction to quantum mechanics* (2018). Cohen-Tannoudji, Diu, and Laloë, *Mécanique quantique* (1991).

[84] Rovelli, "The Relational Interpretation" (2022).

Once quantum theory is learned in this manner, it is hard to view it otherwise. But this is not the way quantum theory was first conceived in 1925 by Born and his pupils, or by Bohr and his, or by Paul Dirac. It is definitely not the only way quantum mechanics can be understood.

The alternative is to understand quantum mechanics as a theory describing discrete facts, connected by probabilistic processes happening when a system interacts with something else. Depending on the interpretation, this "something else" can be the experimenter, the apparatus, the classical world, the subject of QBism or the agent in the pragmatist interpretation. In the relational interpretation, in particular, physical facts are always relative to *two* interacting systems, which in the special case of a laboratory measurement are the measured system and the measuring apparatus. The quantum state $\psi$ plays no fundamental role. It is only a way to code information about some past facts, and a tool for computing probabilities of other facts from those already known.

The understanding of quantum theory offered by these interpretations can be seen as a return to the ideas on which the theory was developed in Göttingen. From the very beginning, under Born's leadership and under the spell of Niels Bohr, quantum theory was seen in Göttingen as a far more radical step than the simple discovery of undulatory properties of matter and radical conceptual departures from classical physics were recognized to be central.

## 3.2 The core ideas of quantum physics

In the fact-based interpretations mentioned above, QM's core ideas are discreteness, probability, contextuality (relationality), and non-commutativity. As in Heisenberg's insight, these interpretations provide the dynamics of the quantum theory via the same equations as in the classical limit, the only specific quantum equation being (1). The following paragraphs illustrate these ideas and their roots in MM.

### *Discreteness rather than continuity*

Born regarded Niels Bohr's discrete "quantum leaps" as the central ingredient of the new physics. A quantum leap is a jump, incompatible with continuity. Born considered the possibility of generalizing this discovery by "replacing differential operations by a difference calculus with the Planck constant".[85] Schrödinger's WM, on the other hand, seemed to have found a way to tame such discontinuities, and this was a hope that Schrödinger and his followers nourished. But nobody better that Schrödinger himself recognized the failure of this hope. Here is what he said years later: "There was a moment when the creators of WM [that is, himself] nurtured the illusion of having eliminated the discontinuities in quantum theory. But the discontinuities eliminated from the equations of the theory reappear the moment the theory is confronted with what we observe"[86].

### *Probability*

The BKS theory that Bohr had developed with Kramers and Slater to compute the amplitudes of atomic transitions assumed fundamental probability. BKS theory turned out to be wrong; contrary to experimental evidence it does not conserve energy in individual processes, which is why Heisenberg was particularly concerned with energy conservation in his Helgoland calculations.[87]  But the main idea that quantum phenomena might have to be directly understood probabilistically was in the air in the early 1920s. Probabilities had been used in atomic physics by Einstein, but generally understood epistemically, namely as expressing our ignorance of the details of well-defined physical situations. Born's group, and also Bohr's, considered probabilities to be an essential part of the new theory.

---

[85] Born, *My Life and My Views,* 34 (1968).
[86] Schrödinger, *Nature and the Greeks and Science and Humanism* (1996).

[87] Aitchison et al., "Understanding Heisenberg's 'magical' paper of July 1925: A new look at the calculational details", illustrates Heisenberg's concern with energy conservation to avoid the problems of BKS theory.

The combination of the discreteness of the quantum leaps and probability imply that the central objects of quantum mechanics are transition probabilities. Mathematically, these are determined by the matrix elements of the evolution operator in the bases determined by the initial and final interactions in a process. For instance, the probability for a jump from a position $x$ to a position $y$ in a time $t$ is determined by (the modulus square of) the matrix element $<x|U(t)|y>$ in the position basis, where $U=exp\{iHt/\hbar\}$. In Feynman's sum-over-histories mathematical formalism, these can be expressed as functional integrals. In MM, they can be derived immediately from the main ingredient of the formalism: matrix elements.

This is explicit in the foundational 1925 paper by Born and Jordan, where quantum theory is formulated in terms of transition amplitudes that determine *the probability* of quantum transitions. Indeed, the paper states explicitly that *the modulus square* of a matrix entry $q_{nm}$ gives the probability of the transitions $n \rightarrow m$.[88] Historical confusion about this point might have been favored by the fact that the English version of the paper published in the influential volume *Sources of Quantum Mechanics* by B.V. van der Waerden is incomplete: the last section of the paper, where transition amplitudes are discussed, is missing in the version published in the book.

Here is how Born himself recollects the origin of this fundamental idea: "We (Jordan and Born) were struck by the fact that the 'transition quantities' appearing in our formulae always corresponded to the square of vibration amplitudes in classical theory. So it seemed likely that the notion of 'transition amplitudes' could be formed. We discussed this idea in our daily meetings, in which Heisenberg often took part, and I suggested that the amplitudes might be the central quantities and be handled by some kind of symbolic manipulation."[89] From the perspective of the interpretations we are considering, this is the core idea needed for formulating quantum dynamics Born was clear in 1925 that the new, future mechanics should yield probabilities for quantum jumps, a fact that makes the motivation of his Nobel Prize (having discovered the probabilistic meaning *of $\psi$*) singularly inappropriate.

---

[88] "Nach Heisenberg sollen die Quadrate der Absolutwerte |q(nm)|² der Elemente von **q** für den Fall, daß **q** kartesische Koordinate ist, maßgebend für die Sprungwahrscheinlichkeiten sein." Born, Jordan, "Zur Quantenmechanik" (1925), 883. This is correct in first order perturbation theory.

A general comment on the role of probability in quantum theory is important in this regard. In some interpretations, like Many Worlds and Bohmian mechanics, quantum probability is epistemic in the sense that a deterministic underlying dynamic is assumed. But the theory remains probabilistic because even in these interpretations the future can be predicted only probabilistically. In Many Worlds, probability comes in through the indexical uncertainty about "us in the future"; we cannot predict in which branch of the universal wave function we ourselves will happen to be; in Hidden Variable theories like Bohmian mechanics, some variables are indeed hidden, preventing complete predictability. Therefore QM is probabilistic *irrespective of the interpretation*.

This last proposition is perhaps not sufficiently emphasized in the literature. The limitation of predictability is irreducible for all interpretations empirically equivalent to the standard theory. It is a radical step away from the classical theory. Even someone firmly believing in 'deterministic' Many Worlds or Bohmian mechanics cannot predict whether a spin-$\frac{1}{2}$ particle will move up or down in a Stern-Gerlach apparatus. Probability is a fundamental feature of *any* theory empirically equivalent to standard quantum mechanics. As Max Born pointed out in his Nobel acceptance lecture, whether an underlying deterministic theory exists or not is in a sense irrelevant, as long as our best available physical theory predicts only probabilities.

### Relationality

The third aspect of quantum mechanics as it was conceived in Göttingen is the most radical and controversial. But it is also the most important. It is the idea of eliminating unobservable quantities from the formulation of the theory. The idea, stated at the opening of the

---

[89] Van der Waerden, *Sources*, 21.

Heisenberg's fundamental Helgoland paper, was in the air at the time and served as a common assumption in Born's Göttingen research group. Its roots can be connected to Einstein's early operationalism and, in the general philosophical atmosphere of the times, attributable to the widespread influence of Ernst Mach[90]. Eliminating electron orbits altogether was an explicit goal of the research programs of Born, Pauli, Kramers, and Heisenberg before the invention of MM.

These men wanted to replace an evolution continuous in spacetime with a set of quantum jumps between manifestations of the system. The amplitudes of these jumps would be the fundamental quantities of QM. As pointed out in § 2, in his 1924 lectures on *Atomdynamik*, Born, following Einstein's hint, had been bold enough to assimilate quantum jumps to radioactive decay, for which no cause could be assigned.

The paper by Born, Jordan, and Heisenberg, states explicitly that the familiar concepts of space and time may not be adequate for describing the motion of electrons [91], an idea later to become profoundly influential in the context of quantum gravity[92].

*Non-commutativity and the fundamental equation*

The only Nobel prize "for the discovery of quantum mechanics" was awarded to Werner Heisenberg for his Helgoland paper. There is no doubt that this paper played a pivotal role in 1925. But it is also true that the paper itself is a mess: to a modern reader it looks confused, incomplete, acrobatic, hard to comprehend.

The originality of the step taken by the 23-year-old Heisenberg was not in the use of matrices (in the sense of tables of numbers); these were implicit in the theory of the time, since the labelling of spectral lines required two integers. Nor did the originality lie in expressing the theory solely in terms of observable quantities; as discussed above, this as well was a position held by many in and out of Göttingen. The game-changing step in the paper was the idea of altering the kinematics —rather than the dynamics— of classical theory by working with matrices as directly representing physical quantities. In the spirit of Bohr's Correspondence principle, Heisenberg found himself in need of the square of one such quantity and, using indirect arguments about the composition of frequencies, he guessed that the matrix $A_{mn}$ corresponding to the square of a quantity represented by the matrix $B_{mn}$ should be

$$A_{mn} = \sum_k B_{mk} B_{kn}. \tag{2}$$

---

[90] Rovelli, *Helgoland* (2021).
[91] "da ja die Elektronenbewegungen nicht in den uns geläufigen Begriffen von Raum und Zeit beschrieben werden können." In Heisenberg "Über quantentheoretische Umdeutung kinematischer und mechanischer Beziehungen" (1925).
[92] Rovelli, *Quantum Gravity* (2004).

Today we recognize that this is matrix multiplication in linear algebra. Heisenberg did not know linear algebra, which was not common knowledge among physicists at the time, and the Correspondence Principle gave no indication that physical quantities were related to linear operations. In the paper, the above equation is buried within a foggy notation and used as an *escamotage*, a sleight of hand, for managing an acrobatic calculation "by a clever analogy"[93]).

The essential novelty about the calculus defined by equation (2) is that it is non-commutative. Born extracted this essential point from Heisenberg's hint, recognized that the sum is matrix multiplication, and guessed that the matrices used by Heisenberg satisfied equation (1). He was able to prove it with the help of Jordan in the 1925 paper. Since the essential novelty of MM is its non-commutative algebra, it is scarcely an exaggeration to say that equation (1) is the only new equation in quantum theory. All quantum phenomena follow from it. This is because the quantum equations are the same as the classical equations (of the classical limit of the theory), with the difference that the variables are non-commutative and satisfy (1). Mathematically, the discovery of quantum theory is the discovery that canonically conjugate physical variables satisfy (1).

<p style="text-align:center">***</p>

Born, Jordan, and Heisenberg completed the formulation of what today we call quantum mechanics in a surprisingly short time. Their "three men" paper is astonishing in containing a complete formulation of quantum mechanics, including the full dynamics, perturbation theory, systems with time dependent Hamiltonian, systems with many degrees of freedom, degenerate systems, continuous spectra, angular momentum, selection rules, the Zeeman effect, and the statistics of quantum field theory, with the derivation of Planck's formula! It is hard to believe that this was completed by November 16, 1925, but that was the case.

---

[93] "durch eine geistreiche Korrespondenzbetrachtung", Born, Jordan, "Zur Quantenmechanik", Introduction (1925).

As pointed out for instance by Mara Beller[94], MM, as defined in 1925, had technical limitations that made its calculations hard for the physicists unacquainted with linear algebra. Perturbation expansion techniques, transformation theory, and other tools to extract physics from were needed. Several of these techniques developed under the influence of Wave Mechanics. All this is true, but it is also true that Quantum Theory was completely defined in 1925 before Schrödinger's work. We can distinguish a theory defined "in principle" from the gradual development of techniques for handling it, extracting information from it, giving mathematical rigor to it, and interpreting it in the most effective manner, which fully build it up "in practice". Thus General Relativity is fully defined by the Einstein field equations, Electromagnetism by the Maxwell equations, and Newtonian gravity by the equations in the *Principia*, although vast developments of their original formulations were needed to exploit them. In this sense, Quantum Mechanics is complete in the 1925 papers of the Gottingen group. Given a theory with canonical position and momentum variables, equation (1) and the Hamiltonian as a function of these variables define the mathematics of the theory completely.

Beller also points out that the founders of MM tempered the radicality of their ideas during intense discussions in the immediate aftermath of the publication of WM. This is certainly true, but many current interpretations of Quantum Mechanics suggest that this partial retreat was not a step towards clarity, but rather a stumbling block on the way to clarification. The radical Göttingen idea that atomic phenomena cannot be conceived as located in space and time was indeed severely challenged by the spatiotemporal visualizability of the Schrödinger function. But this visualizability in spacetime is misleading since it does not extend beyond the single particle theory, let alone to quantum field theory or quantum gravity. Features of MM such as lacking a fundamental notion of state, a continuous evolution in time, and a spatio-temporal interpretation, which were viewed as difficulties at the time and are still viewed as problematic by some physicists today, are actually now counted as virtues of the theory by others. For instance, not only are modern views of QM like Q-bism or the Relational Interpretations consistent with the original radical Göttingen ideas, but even interpretations like Many Worlds have abandoned spacetime as the proper arena for quantum physics. Historically, WM turned out to be a much handier formalism for addressing helium, collisions, and so on, but, shortly after these successes, Quantum Field Theory developed in a spirit closer to MM than WM. The particle number basis is commonly employed, while the "wave functional of the field" representation and the Schrödinger picture of the evolution are possible, but rarely employed.

---

[94] Beller ,"Born Probabilistic interpretation: a case study of 'concept in flux" (1990).

The name "Heisenberg" appears twenty-four times in the fundamental Born-Jordan paper. The introduction states that "Heisenberg has expressed the physical concepts that guided him so clearly that any extended remark would seem superfluous." This is an astonishing and totally false statement: Heisenberg's paper is a patchwork. This generosity is a beautiful aspect of Born's reserved and shy personality, quite different from that of many other flamboyant characters in the quantum saga. There were other scientists not focused on recognition; think of Lemaître, for instance, who erased a key paragraph from the English translation of his own paper, leaving all the credit for the discovery of the expansion of the universe to Hubble[95]. But they are rare. Many scientists will go far to claim credit owing to others, sometimes only for having corrected a detail, or filled in a little hole. This behavior is encouraged by the romantic idea of the discoverer. And that idea abets a misreading of the opening pages of the papers by the Göttingen collaborators. Their praise of the Helgoland breakthrough easily gives the impression to anyone who had not read the Helgoland paper that Born and Jordan's work was an appendix or pendant to Heisenberg's: but a study of all the papers makes it clear that quantum theory was developed in the Born-Jordan paper and in the "three men" paper, with just a hint from the Helgoland paper.

---

[95] Livio, "Lost in translation: Mystery of the missing text solved" (2011).

Some physicists in a similar situation would have claimed the full merit for themselves and dismissively mentioned the inspiration from Heisenberg's paper in a footnote. Born's generosity may have cost him recognition for the paternity of the new physics.

There is one fact that diminishes the centrality of MM: Paul Dirac, also inspired by the Helgoland paper, got to its main equation independently. Dirac's 1925 paper[96] does not have the complexity and the full development of the theory in the three men paper, but has the basic equations, in a form which is extremely terse and clean. Dirac too recognized that the key ingredient was non-commutativity, treated it abstractly with non-commutative q-numbers, and recognized the connection with classical Poisson algebra.

Non-commutativity and (1) define quantum theory entirely: the eigenvalues of the q-numbers, which can be defined algebraically,[97] give the possible values of a physical quantity for any (ordered) function $A(q,p)$ of $p$ and $q$. Transition amplitudes between eigenstates corresponding to these eigenvalues can be defined similarly. The time evolution of any quantity is given by the commutator with the Hamiltonian. This terse algebraic formulation of quantum theory, which is empirically equivalent to wave functions, Hilbert spaces, path integrals, and so on, was sufficient for Pauli to compute the spectrum of the hydrogen.

This way of conceiving quantum theory is very close to the understanding of the theory provided by modern interpretations that deemphasize the ontological role of the quantum state. In these, the quantum state $\psi$ plays only an auxiliary role, similar to the role played by the Hamilton-Jacobi function $S$ in classical mechanics (which is in fact directly related to its classical limit: $\psi \sim \exp iS/\hbar$): a tool for calculations and a way to code information about what is known of the system's past manifestations.

When quantum physics is formulated in these terms, the measurement problem takes a different form from the question of what causes the collapse of $\psi$. The ontology of the theory is not based on $\psi$. It is directly given by the observable facts, the discrete manifestation of the quantum systems, that are described by eigenvalues of q-numbers. These are values that physical quantities can take in the moment the quantum system interacts with another system.

---

[96] Dirac, "The fundamental equations of quantum mechanics" (1925).
[97] The numbers $a$ for which $(A - a\mathbb{1})$ has no inverse.

In different interpretations, these discrete elementary interactions are interpreted differently. The Copenhagen interpretation assumes the existence of a classical world and the relevant interactions are discrete physical encounters between the quantum system and the classical world. QBism starts from the idea of an agent that acquires information about the external world and the interactions are those between the agent and the world. In Relational Quantum Mechanics, essentially any interaction can play this role, but the values that variables take at interactions are relational: they are relative to the interacting system. In all these cases, quantum mechanics is interpreted as a radical conceptual novelty with respect to classical mechanics: the best description we can give of the world is discrete and probabilistic; it concerns the way systems manifest themselves, and not how systems *are* between their manifestations. All these views are very close to and have their root in quantum mechanics as understood by Max Born and Niels Bohr.

### 3.3 WM and the probabilistic interpretation of $\psi$

Schrödinger obfuscated the momentous achievement of the two main papers of Born and his collaborators by rederiving the spectrum of the hydrogen atom using differential equations rather than algebra. Differential equations often enable an easier computation of eigenvalues than algebraic methods.

Schrödinger's result had a sudden success for several reasons. Linear algebra was not taught in most universities, while differential equations were used by all theoretical physicists. As a way to calculate, the Schrödinger equation proved to be effective. More importantly, Schrödinger's wave function and its simple "visualizability" immediately captured attention because it appeared to avoid the extreme radicality of the Göttingen ideas. WM raised the hope that a radical conceptual revolution could be avoided. The fact that a century later we are still debating what precisely we have learned about nature with the discovery of this theory clearly shows that this hope was disappointed.

In the early days of the theory, the hope that it could be reduced to the discovery of a wave nature of matter processes was high. Take the letter written by Einstein in which he considers the idea of recommending Heisenberg-Bohr-Jordan for the Nobel Prize. The letter opens making clear that the new advances in quantum physics are "the most important [...] work in physics" but characterizes it as "knowledge of the wave nature of matter". His sympathy, as well known, went towards the de Broglie-Schrödinger version of the theory, which he still hoped, unsuccessfully, to turn into a regular field theory. He described the Heisenberg-Born-Jordan achievements as "only theoretical", which is curious: it is true that electron diffraction could be seen as direct support for de Broglie's matter waves, but there are more connections to open questions in atomic theory in the three men paper than in all the de Broglie and Schrödinger papers on the topic.

For many physicists, Schrödinger's formulation "came as a great relief, now we did not any longer have to learn the strange mathematics of matrices"[98], but the relief was illusory. Heisenberg reacted very negatively and vocally to the success of WM, arguing —with good reasons— that it was confusing the issue, not clarifying it. He lost the political battle because after all few were interested in conceptual clarity, the practical utility of the differential equation was undeniable, and Bohr mediated between the bickering parties, waving his hands over a 'wave particle duality'. Characteristically for his low-key personality, Born kept himself on the sidelines of the fierce debates. But his role was again crucial with the 1926 paper[99] where he introduced the probabilistic interpretation of Schrödiger's wave function. This paper deserves a closer look in this context.

Interestingly, the paper does not derive the fact that the (modulus square) of the value of the wave function at a point, $\psi(x)$, gives the probability density for the particle to be in $x$ (this was later observed by Pauli, as mentioned above). Rather, the paper interprets the coefficients of the expansion of the wave function $\psi(x)$ into a basis as the quantities that determine the probability of a quantum transition. This can be immediately recognized as

---

the application to WM of the probabilistic logic of the MM of the Born-Jordan paper. And this of course is the right general interpretation of probability, because in standard QM $\psi(x)$ is not the probability density for the particle to be in $x$: it is the probability density that the system be found in $x$ when a measurement with something determining its position is made.

Why did Born resort to WM to state something that is also true in his MM? After all, the mathematical equivalence between WM and MM was immediately recognized by Schrödinger. The reason is stated in the opening of the paper: MM, in 1926, had difficulty in treating scattering. The theory was first formulated in the energy eigenbasis and it took Jordan's[100] and Dirac's[101] transformation theory, and von Neuman's formalization, to fully clarify the basis independence of quantum mechanics. The problem was to identify the correct basis to represent incoming and outgoing particles in a scattering process, and this was complicated by the continuous spectrum of the momentum. In WM, which is quantum theory in the position eigenbasis and development expressed in terms of quantum states rather than physical variables, incoming and outgoing states with given momentum can be intuitively recognized as plane waves, as in de Broglie's picture. This is what Born did in 1926. He resorted to WM because it provided an intuitive shortcut. Only with the development of transformation theory (which reinforced and axiomatized his statistical approach) did Born reach his final stand: together with Heisenberg and Jordan, and in contrast to his initial response, Born came to see Schrödinger's wave mechanics as no more than a mathematical appendix to matrix mechanics. In his words: "Schrödinger's achievement reduces to something purely mathematical".[102]  On this, see also Wessel[103] and Beller[104].

This ended up earning him the Nobel prize, but ironically obscured the relevance of *his* early major ideas by implying that WM was essential for this step. Ironically, today quantum

---

scattering is largely treated in particle physics using quantum field theory, where virtually nobody uses WM.

### 3.4 A counterfactual story

To illustrate the theoretical point about the relation between WM and MM, let us ask what might have plausibly happened if WM had not appeared on the scene in 1925. Any counterfactual history is arbitrary, given the large role of serendipity in the flow of life.   But from a theoretical perspective it does make sense to ask whether some scientific idea *could* have evolved in the absence of an ingredient that played a historical role, because in this way theoretical arguments regarding the need of this ingredient for the internal coherence of a theory can be evaluated.

WM would not have appeared on the scene in 1925 if its prime mover, Louis de Broglie, had followed a career more consistent with his family's aristocratic position than the cultivation of physics.  This would have left Schrödinger with no waves to navigate.

No doubt someone would have discovered the Schrödinger equation and the $\psi$ state had Schrödinger not done so. In this counterfactual history, however, this would have happened a few years later, without de Broglie's intuitive idea that the particles "are" waves, and after the power of quantum theory in its matrix formulation had been fully displayed. It would have happened after Jordan and Dirac had recognized the basis independence of Göttingen's matrices and worked out transformation theory, and the mathematicians had clarified the properties of non-commutative algebras and the $q$-numbers with continuous spectrum.

A little-known paper by Carl Eckart[105], which appeared in 1926, shows how WM could have been found from MM without passing through Schrödinger. Eckart discusses the equivalence of WM and MM, as Schrödinger had independently done shortly earlier; but, unlike Schrödinger, he does so starting from MM.   The wave function $\psi$ that satisfies the Schrödinger equation appears as the generalization of the "numerical quantity" *exp*[$i$Et/ℏ] that Born and Wiener had used shortly earlier in their pioneering work on operators[106]. Searching for a convenient method for solving the eigenvalue problem for the energy would have led to considering the problem in the resulting position eigenbasis, and therefore to the time independent Schrödinger equation.

---

[105] Eckart "Operator Calculus and the Solution of the Equations of Quantum Dynamics" (1926).

The same fruitful transfer of techniques and interpretations from WM to MM and vice versa, which made the development of transformation theory possible and overcame the early technical shortcomings of MM, would have happened without charging $\psi$ with ontological baggage. Everyone would have perceived the resulting WM as Heisenberg did in fact view WM: as a useful tool for calculation, but merely a tool.

The connection between the Schrödinger equation and the Hamilton-Jacoby equation, which Schrödinger used as his inspiration in seeking a "WM" by inverting a presumed eikonal approximation[107], would have simply come about as the quantum version of the analogous connection in classical mechanics, where the Hamilton-Jacobi theory is interpreted as a fancy way for doing calculations, devoid of any direct ontological weight. That is, since Hamilton-Jacobi function emerges from the semiclassical approximation of Schrödinger's wave function, it would have been natural to ascribe to the function the same interpretation as the Hamilton-Jacobi function: a mere calculational tool, not the description of a real entity. After all, the dynamics of a *classical* particle can be described in terms of a function on spacetime, the Hamilton-Jacobi function, without pushing anybody to an undulatory interpretation of classical mechanics.

Had this been the history, the mathematical development of quantum mechanics and its predictive power would have converged to the same form as today, but the interpretation problem would have been markedly different. Electron diffraction would have been discovered, since it is predicted by the transition amplitudes of quantum mechanics, but perhaps later, and not by the son (G. P. Thomson) of the discoverer (J. J. Thomson) of the electron.

The question would have never been how does $\psi$ "collapse" during a measurement. Rather, the question would have been what determines the occurrence of the quantum events whose probabilistic relations are described by the transition amplitudes of the theory. Bohr could have emphasized, as he did in fact, the contextuality of every quantum event, but he might well have articulated this more generally as relative to any possible measurement, and not as complementarity between "particle" and "wave" pictures. The standard textbook interpretation (contextuality, measurements, eigenvalues and eigenstates, probabilities, Heisenberg principle...) would have ended up in the same place as it has, but not soiled by the intuition that particles are waves and by the associated $\psi$-ontology. Einstein would not have hesitated between suggesting de Broglie-Schrödinger versus Heisenberg-Born-Jordan for the Nobel and would not have written that "the most important, certain, and unawarded work in physics is the knowledge of the *wave* nature of material processes." Heisenberg's discovery of the uncertainty relations would have solidified the interpretation and the associated contextuality. Einstein's concerns for the lack of a fully observer-independent realistic picture would have led him towards considering the EPR correlations, and Bell's strong realist intuition would have led him to the equalities that would follow if the probabilistic correlations were underpinned by a local hidden variable theory. The full ensuing discussion, all the way to the recent Nobel for the experimental confirmation that the Bell equations are violated, would have been the same.

But something would have been quite different. The intuition that the "real stuff" underpinning the quantum phenomena is a continuous wave, or a continuously evolving state, would not have taken hold. Everett could have equally remarked that the quantum state we attribute to a system is always relative to a second system (opening the way to relational and perspectival interpretations), but this crucial observation would not have led to the hypothesis of a self-standing overall wave function as the basis of the quantum ontology, as is assumed in the Many Worlds and related interpretations. Similarly, physical collapse theories trying to modify the Schrödinger equation to provide a hypothetical dynamical mechanism underlying measurements events would have looked bizarre, given the marginal role of the Schrödinger equation. Quantum gravity, where the "Schrödinger picture" (evolving states) is not viable and only the "Heisenberg picture" is[108], would have been easier to develop.

As it happened, however, WM did come along in time to enrich, but confuse, the interpretation of QM with a persistent psi-ontology invented by de Broglie and Schrödinger and absent from Born's MM.

## 4. Conclusions

We have brought out some of the accidents of circumstance and personality that explain the oddities of Born's Nobel Prize. In the process, we have presented the relation between MM and WM in contrasting and complementary ways. Although it was not our purpose to award Max Born more credit than he is usually given, our analysis unavoidably led us to do so. Born introduced the term "quantum mechanics" for the new theory, and some suggestions for achieving it, in 1924, a year before its invention[109]. With Jordan, he found its fundamental equation $\mathbf{pq} - \mathbf{qp} = \frac{h}{2\pi i}\mathbf{1}$ and recognized that the essential novelty of the new mathematics in the non-commutativity of its physical variables. With Jordan and Heisenberg, he wrote the first paper developing an almost complete QM. He identified the key ingredient of the theory as transition probabilities and pointed Heisenberg in the right direction: in Heisenberg's words, "It was the peculiar spirit of Göttingen, Born's faith that nothing short of a new self-consistent quantum mechanics was acceptable as the goal in fundamental research that enabled [my] ideas to come to full fruition."[110]. Inspired by Einstein, Bohr, and Kramers, Born recognized the role of probabilities in the theory and understood how to compute them. His caution and self-doubt, his generosity and unassertiveness, left the scene

---

and credit to his flamboyant collaborators. Yet, the way his papers introduced the theory is in tune with some contemporary perspectives.

We have also attempted to elucidate central points by imagining a counterfactual history, in which WM would not have appeared in 1926. That was not how it happened. The way it happened was marked and sometimes marred by the culture and personalities of the actors, the character and traditions of their science, and the wider world in which their work developed. The regard and appreciation of the scientific community for Bohr's contributions and MM evolved as the generations of quantum physicists and the historians and philosophers interested in their work have come and gone. Our views do not occupy a privileged position in this flux: they are but one (or rather two!) of the many ways to appreciate the discoveries of QM and the inventions from which they sprang.

## 4. Ad Pleniorem Scientiam

*CR: Well, what do you think of my counterfactual history?*

*JH: It does very well what you want it to do and could be elaborated if anyone thought it necessary. For example, the detection of electron diffraction by Davisson and Germer and by George Thomson did not have to pass through de Broglie. Davisson and Germer had discovered an instance of it by themselves and Thomson prepared the necessary apparatus not to test wave mechanics but to develop one of his father's most outmoded ideas. And a wave-like property of an electron beam might have been inferred from the Ramsauer effect and Lord Rayleigh's discovery that light of long wavelength passes easily through an atmosphere of small particles.*

*CR: I would have thought that as a historian you would not want to be entangled in speculating on a development that did not take place.*

*JH: It is not my favorite form of historical writing, but it has the merit of drawing attention to contingencies in historians' reconstructions. And so, your counterfactual history inspires me to concoct another. Let us suppose that Heisenberg on that fatal night in Helgoland ran into an intelligent attractive girl and gave up physics for many nights; that he did not return to*

Göttingen with his breakthrough paper; and that he spent the next six months hiking with his pathfinders and improving his Danish. No Helgoland paper, no matrix mechanics.

CR: A few months later, in the Alps, a woman, certainly intelligent and attractive, did not interfere negatively with Schrödinger's physics...

JH: Schrödinger was more experienced in these matters.

CR: Anyway, what follows from this hypothetical distraction of Heisenberg?

JH: Schrödinger's waves would have excited Heisenberg's competitive instincts and he would have shown how to solve old problems and approach new ones using the new formulation. He would have treated ortho- and para-helium and discovered resonance effects and exchange forces and Born would have proposed his probabilistic interpretation of the wave function, just as they did in fact. And to add a fact, these were the accomplishments that Dirac's teacher, Ralph Fowler, mentioned in 1927 in favor of WM when deliberating whether it or MM was the more fundamental.

CR: And I suppose that you are about to conclude that since the reciprocal relations of $\Delta p$ and $\Delta q$ follow from the wave concept, physicists would have found them without introducing the pseudo problems of "disturbing the system" associated with Heisenberg's presentation of his "uncertainties."

JH: Precisely. "Disturbing the system" implies the pre-measurement existence of a definite state and some sort of ontology. Without being burdened by the complications of a non-existent MM, Heisenberg and Born would have shared a Nobel prize for their profound elaborations of WM.

CR: Clever. But the problems would have shown up very soon: the clash between Schrödinger's initial realism about his wave and the probabilistic interpretation would have been even more devastating.

JH: I do not claim that my counterfactual history is more plausible than yours; I say only that it brings out other contingencies and perhaps says something useful about Heisenberg's first formulation of his uncertainty principle.

CR: Quite. And it has the merit of suggesting a more correct reading of Einstein's position at the time of EPR; he was then no longer trying to undermine the uncertainty relations but rather trying to call attention to entanglement and its bizarre consequences. I suspect he got to EPR by realizing that Bohr's alleged "solution" of the famous "box of light" (or "photon in the box") puzzle is nonsense. That would have helped identify the perturbing role of psi-ontology earlier.

JH: It seems to me that avoidance of psi-ontology -- by which I mean the ascription of some reality to the theoretical constructions of QM -- may not be so easy as your characterization of the results of measurement as "manifestations" suggests.

CR: How so?

JH: What is it that manifests? Although physics may not be able to say what happens to "it" or "the system" between measurements, what allows us to assume that the same "it" manifests itself in both instances? And does it help any to call the "it" a system?

CR: Don't we do the same in classical mechanics? We posit an entity to fill in for its manifestations. There is nothing wrong with this, as long as it works. Problems start, I believe, if we remain attached to the details of a theory or hypothesis when it does not work anymore. It is the great lesson of Ernst Mach, that has inspired both revolutions of modern physics.

JH: But did it? Einstein thought that Mach's version of positivism had inspired special relativity until he discovered that Mach rejected it. And Max Planck attacked Mach's doctrine as a clear and present menace to physics. I wonder whether anyone has ever discovered anything while thinking positivistically. When you think about physical problems, does your imagination supply any picture of the situation between quantum jumps?

CR: In his great play Copenhagen, Michael Frayn has Heisenberg say that he got the basic intuition about quantum uncertainty by thinking about a man at night appearing and disappearing under the light of a lamp: we have no trouble, and get into no trouble, supposing that he continues to exist between our glimpses of him. The man is a heavy object, he can move and endure without quantum effects. But what do we know about electrons?

JH: Whatever we please. They are at our disposal, we fashion them to assist our planning and understanding of experiments. To return to my question, how, if we can conceive of no *connection between manifestations, are we to understand how nature brings about the statistical results that QM enables us to calculate?*

*CR: One could just as well ask how nature forces the electric field to satisfy the Maxwell equations? Or a stone to fall as it does? Newton ended his great book with an answer to such questions that I believe opened the way to modern science:* hypotheses non fingo. *Science advances by finding questions that can be answered. Then we adapt our intuition to the phenomena, not the other way around.*

*JH: But Maxwell and Newton did not discover their equations by eschewing hypotheses; it seems to me not entirely honest to declare, with the equations in hand, that we should not ask the sorts of questions that led the discoverers to them. I am reminded of Bohr's teaching that when contradictory concepts are needed to account fully for experience, we must consider them as complementary.*

*CR: There may well be a complementarity between making physics and philosophizing about it. That makes an appropriate ending to our discussion. Born followed Bohr's lead from Correspondence to Complementarity, and for good reason. Bohr's manner of thought, his willingness to entertain problems of principle, and his insights, at once vague and penetrating, perfectly suited the first explorations of the tantalizing, elusive quantum world.*

## Acknowledgments

This research was made possible thanks to the project on the Quantum Information Structure of Spacetime (QISS) supported by the JFT grant 61466.